\documentclass[aps,prl,twocolumn]{revtex4}
\usepackage{amssymb}

\usepackage{graphicx}

\begin{document}

\title{Determining the in-plane Fermi surface topology in underdoped high $T_{\rm c}$ superconductors using angle-dependent magnetic quantum oscillations}
\author{N.~Harrison and R.~D.~McDonald}
\affiliation{National High Magnetic 
Field Laboratory, Los Alamos National Laboratory, MS E536,
Los Alamos, New Mexico 87545}
\date{\today}

\begin{abstract}
We propose a quantum oscillation experiment by which the rotation of an underdoped YBa$_2$Cu$_3$O$_{6+x}$ sample about two different axes with respect to the orientation of the magnetic field can be used to infer the shape of the in-plane cross-section of corrugated Fermi surface cylinder(s). Deep corrugations in the Fermi surface are expected to give rise to nodes in the quantum oscillation amplitude that depend on the magnitude and orientation of the magnetic induction ${\bf B}$.
Because the symmetry of electron and hole cyclinders within the Brillouin zone are expected to be very different, the topology can provide essential clues as to the broken symmetry responsible for the observed oscillations.
\end{abstract}

\maketitle

The recent discovery of magnetic quantum oscillations in high temperature superconductors provides an entirely new perspective on their electronic structure~\cite{doiron1,yelland1,bangura1,sebastian1,vignolle1}. These experiments provide evidence for a Fermi surface and possible evidence for a broken symmetry groundstate in underdoped samples, yet the precise form of the symmetry remains the subject of debate. A key point of discussion is the carrier sign of the pocket responsible for the largest amplitude quantum oscillations in YBa$_2$Cu$_3$O$_{6+x}$ and YBa$_2$Cu$_4$O$_8$~\cite{doiron1,leboeuf1}. Quantum oscillation experiments cannot directly infer the sign of the carriers, leaving open many possibilities for the interpretation of the negative Hall coefficient and topology of the Fermi surface~\cite{millis1,chen1,hossain1,dimov1,hozoi1,lee1,varma1}. The choice between electron and hole pockets cannot be taken lightly. If the carriers are truly electron-like then this places limitations on the types of theoretical model that can be applied to the underdoped metallic state of high $T_{\rm c}$ superconductors. Given the possible implications for the competing order parameters in YBa$_2$Cu$_3$O$_{6+x}$,  a thorough experimental investigation of the Fermi surface topology is of paramount importance.

\begin{figure}[htbp!]
\centering
\includegraphics[width=0.45\textwidth]{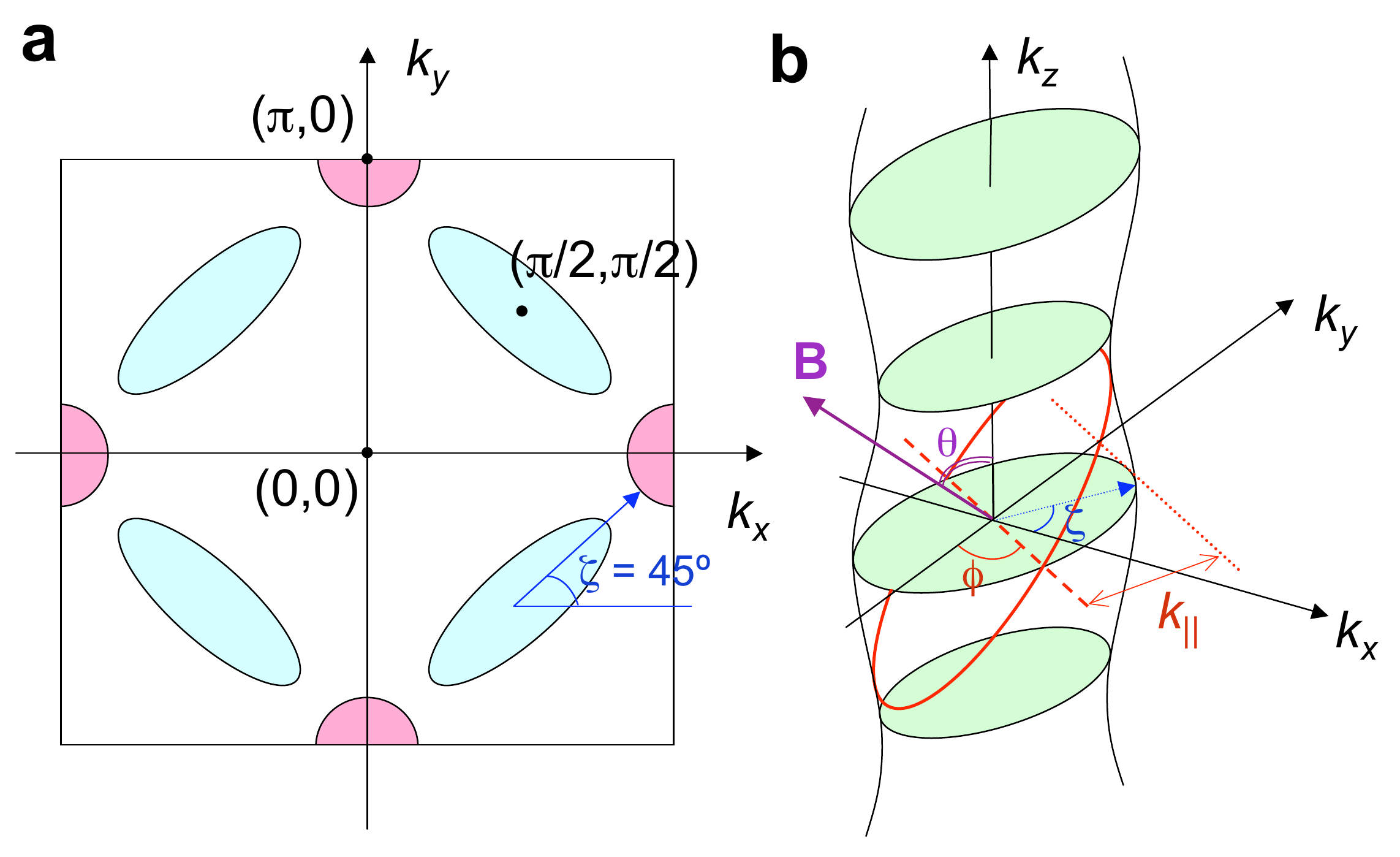}
\caption{{\bf a} Schematic Fermi surface for the simple case of ${\bf Q}=(\pi,\pi)$ density-wave ordering, leading to reconstructed electron (pink) and highly elliptical hole (cyan) pockets. The electron pockets are expected to vanish in strong-coupling models. {\bf b} Schematic warped Fermi surface with an elliptical cross-section with its major axis at an arbitrary orientation $\zeta$ within the $k_x$, $k_y$ plane. The extremal cross sections orthogonal to $k_z$ and ${\bf c}$ are depicted in green. The red curve shows an example orbit when the magnetic induction ${\bf B}$ is rotated by a `zenithal' angle $\theta$ away from the $k_z$ axis. The half caliper width of the in-plane cross-section $k_\|$  depends on the `azimuthal' component $\phi$ of the magnetic field orientation and on $\zeta$.}
\label{schematic}
\end{figure}

In this paper we show that crucial clues as to the nature of the broken symmetry phase that competes with superconductivity may be found by performing quantum oscillation experiments in which the crystalline $c$ axis of the sample is tilted away from axis of the magnetic induction ${\bf B}$ in different directions. The recently proposed deeply corrugated form of the Fermi surface cylinder(s)~\cite{audouard1,sebastian2} implies the existence of beats in the quantum oscillation amplitude as a function of the magnitude and orientation of ${\bf B}$. By tracking the position of nodes in the quantum oscillation amplitude, the shape of the in-plane cross-section of the cylinders can be mapped out in a similar fashion to angle-dependent magnetoresistance oscillations (AMRO)~\cite{nam1}. A distinction between electron and hole pockets can be made because they are predicted to have different symmetry properties within the Brillouin zone (as depicted for the simple case of ${\bf Q}=(\pi,\pi)$ ordering in Fig.~\ref{schematic}a), the details of which may further depend on the precise form of the broken symmetry.

Independent of the broken symmetry responsible for the creation of a given corrugated Fermi surface cylinder, the $k_z$-dependence of the cross-sectional area $A_k$ can be expanded as a series of harmonics. A schematic of a corrugated Fermi surface cylinder is shown in Fig.~\ref{schematic}b. If the in-plane crystal lattice vectors ${\bf a}$ and ${\bf b}$ are orthogonal to the interlayer vector ${\bf c}$, $A_k$ can be considered as the sum of even harmonics
\begin{equation}\label{area}
A_k=\sum_nA_n\cos(nk_zc)
\end{equation}
where $c=|{\bf c}|$. Here, $A_0$ is the mean cross-sectional area of the Fermi surface, $A_{n\geq 1}\approx4\pi m^\ast t_{c,n}/\hbar^2$ represent harmonics of the corrugation and $m^\ast$ is the in-plane effective mass. 
In the presence of a magnetic induction ${\bf B}$, the $k$-space area $A_{\theta,\phi}$ of a given semiclassical orbit (depicted in red in Fig.~\ref{schematic}b) depends on the `zenithal' $\theta$ and `azimuthal' $\phi$ orientation of ${\bf B}$ with respect to the corrugated cylinder (as depicted in Fig.~\ref{schematic}b). The orbit area in such a situation was found by Yamaji~\cite{yamaji1} to have the form
\begin{equation}\label{yamaji}
A_{\theta,\phi}\cos\theta=\sum_nA_n{\rm J}_0(nk_\|c\tan\theta)\cos(nk_{z,0}c).
\end{equation}
Its further dependence on the half caliper width $k_\|$ of the corrugated cylinder within the $k_x$,$k_y$ plane was considered in subsequent works~\cite{nam1}. Here ${\rm J}_0(x)$ is the zeroth order Bessel function, while $k_{z,0}$ is the $k_z$ coordinate of the geometric center of the orbit. If the in-plane cross-section is elliptical in form, as if often approximately the case for small pockets of carriers~\cite{nam1}, then
\begin{equation}\label{caliper}
k_\|=k_{\rm F}\sqrt{\eta\cos^2(\phi-\zeta)+\frac{1}{\eta}\sin^2(\phi-\zeta)},
\end{equation}
where $k_{\rm F}=\sqrt{A_0/\pi}$ is the effective Fermi radius, $\eta$ is the ellipticity of the pocket (or the ratio of the minor axis to the semi-latus rectum, or the ratio bsetween major and minor axes) and $\zeta$ is the orientation of its major axis with respect to $k_x$ within the $k_x$, $k_y$ plane. For $\eta=$~1, this reduces to a perfect circle where $k_\|=k_{\rm F}$, as was originally considered by Yamaji~\cite{yamaji1}. The $k_{z,0}$-dependence of the quantum oscillation frequency is obtained applying Onsager's relation
\begin{equation}\label{onsager}
F_{\theta,\phi}=\bigg(\frac{\hbar}{2\pi e}\bigg)A_{\theta,\phi}
\end{equation}
to Eqn~(\ref{yamaji}).
Finally, the form of the magnetic quantum oscillations is obtained by integrating over $k_{z,0}$. On neglecting oscillations of the chemical potential (assuming that the corrugation is large), the oscillatory magnetization is given by~\cite{shoenberg1}
\begin{equation}\label{magnetization}
M_z=-\frac{1}{\pi}\int^{\pi/c}_{-\pi/c}\sum_{\zeta,p,n}M_p\sin\bigg(2\pi p\bigg(\frac{F_{\theta,\phi}}{B}-\frac{1}{2}\bigg)\bigg){\rm d}k_{z,0}
\end{equation}
where
\begin{equation}\label{prefactors}
M_p=\bigg(\frac{\hbar eN}{pm^\ast}\bigg)\frac{X_p}{\sinh X_p}\exp\bigg(-\frac{p\Gamma}{B}\bigg)\cos\bigg(\frac{\pi pm^\ast g}{m_{\rm e}\cos\theta}\bigg)
\end{equation}
are the conventional amplitude-modifying prefactors and $X_p=2\pi^2pm^\ast k_{\rm B}T/\hbar eB$~\cite{shoenberg1}. Here, $p$ is the quantum oscillation harmonic index. The term containing the hyperbolic sine is the thermal damping factor, the exponential factor accounts for quasiparticle scattering, while the cosine term accounts for Zeeman splitting (or `spin splitting')~\cite{shoenberg1}. 

\begin{figure}[htbp!]
\centering
\includegraphics[width=0.45\textwidth]{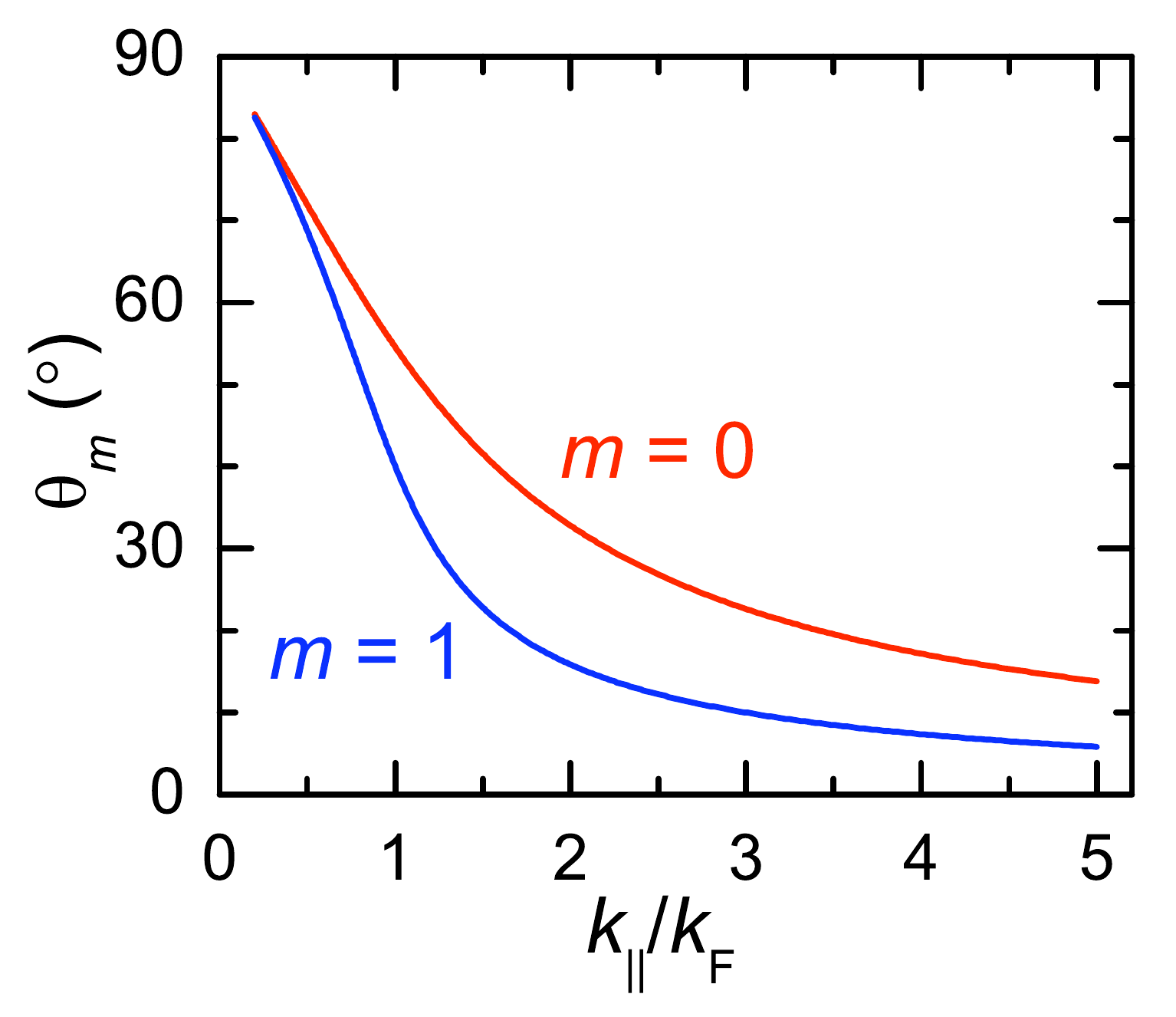}
\caption{A plot of $\cot\theta_m$ versus $k_\|/k_{\rm F}$ according to Eqn (\ref{nodes}) for $m=$~0 and $m=$~1, using $c=$~11.68~\AA~ and $\Delta F/B=$~1 (approximately applicable to YBa$_2$Cu$_3$O$_{6.54}$ at $B=$~40~T). For simplicity, we consider only the nodes occurring when $k_\|c\tan\theta_m<\frac{3\pi}{4}$.}
\label{nodeplot}
\end{figure}

To understand the basic form of the ${\bf B}$-dependent quantum oscillation amplitude, it is instructive to first consider the simpler situation in which the second harmonic of the corrugation $A_{n\geq 2}$ is neglected (along with higher harmonics). In this case, the integral given by Eqn~(\ref{magnetization}) has the simple solution
\begin{eqnarray}\label{magnetization2}
M_z\approx-\frac{1}{\pi}\sum_{\zeta,p}M_p{\rm J}_0\bigg(\frac{2\pi p\Delta F_{\theta,\phi}}{B}\bigg)
\times~~~~~~~~~~~~~~~\nonumber\\\sin\bigg(2\pi p\bigg(\frac{F_0}{B}-\frac{1}{2}\bigg)\bigg),
\end{eqnarray}
where $\Delta F_{\theta,\phi}=\Delta F{\rm J}_0(k_\|c\tan\theta)/\cos\theta$ is the difference in frequency between maxima (`belly') and minima (`neck') frequencies of the corrugated cylinder (with $\Delta F$ being that at $\theta=$~0). In AMRO experiments, one is typically concerned with the `magic' angles at which $\Delta F_{\theta,\phi}$ vanishes~\cite{yamaji1,nam1}$-$ where the electronic structure mimics that of an ideal two-dimensional metal. However, nodes in the quantum oscillation amplitude, which occur for a finite value of $\Delta F_{\theta,\phi}$, can be determined with greater accuracy in quantum oscillation experiments. Since  $\Delta F_{\theta,\phi}>B/8\pi$ in the limit where we consider nodes, Equation (\ref{magnetization2}) can be accurately approximated by the superposition
\begin{eqnarray}\label{magnetization3}
M_z\approx-\frac{1}{\pi^2}\sum_{\zeta,p}M_p\sqrt{\frac{B}{p\Delta F_{\theta,\phi}}}
\times~~~~~~~~~~~~~~~~~\nonumber\\\sin\bigg(2\pi p\bigg(\frac{F_0\mp\Delta F_{\theta,\phi}}{B}-\frac{1}{2}\bigg)\pm\frac{\pi}{4}\bigg)
\end{eqnarray}
of neck $F_{\rm neck}=F_0-\Delta F_{\theta,\phi}$ and belly $F_{\rm belly}=F_0+\Delta F_{\theta,\phi}$ quantum oscillations, giving rise to a distinctive beat pattern. Nodes occur whenever the $F_{\rm neck}$ and $F_{\rm belly}$ oscillations are $\pi$ out of phase, which occurs at a different set of `magic' angles $\theta_m$ given by
\begin{equation}\label{nodes}
{\rm J}_0(k_\|c\tan\theta_m)=\frac{(4m+3)B}{8\Delta F}\cos\theta_m,
\end{equation}
where $m$ is an integer. The monotonic dependence of the $m^{\rm th}$ node $\theta_m$ on $k_\|/k_{\rm F}$ in Fig.~\ref{nodeplot} (in which $m=$~1 and 0 are plotted) implies that its $\phi$-dependent angular position can be used to map the shape of the cylinder cross-section.

\begin{figure}[htbp!]
\centering
\includegraphics[width=0.45\textwidth]{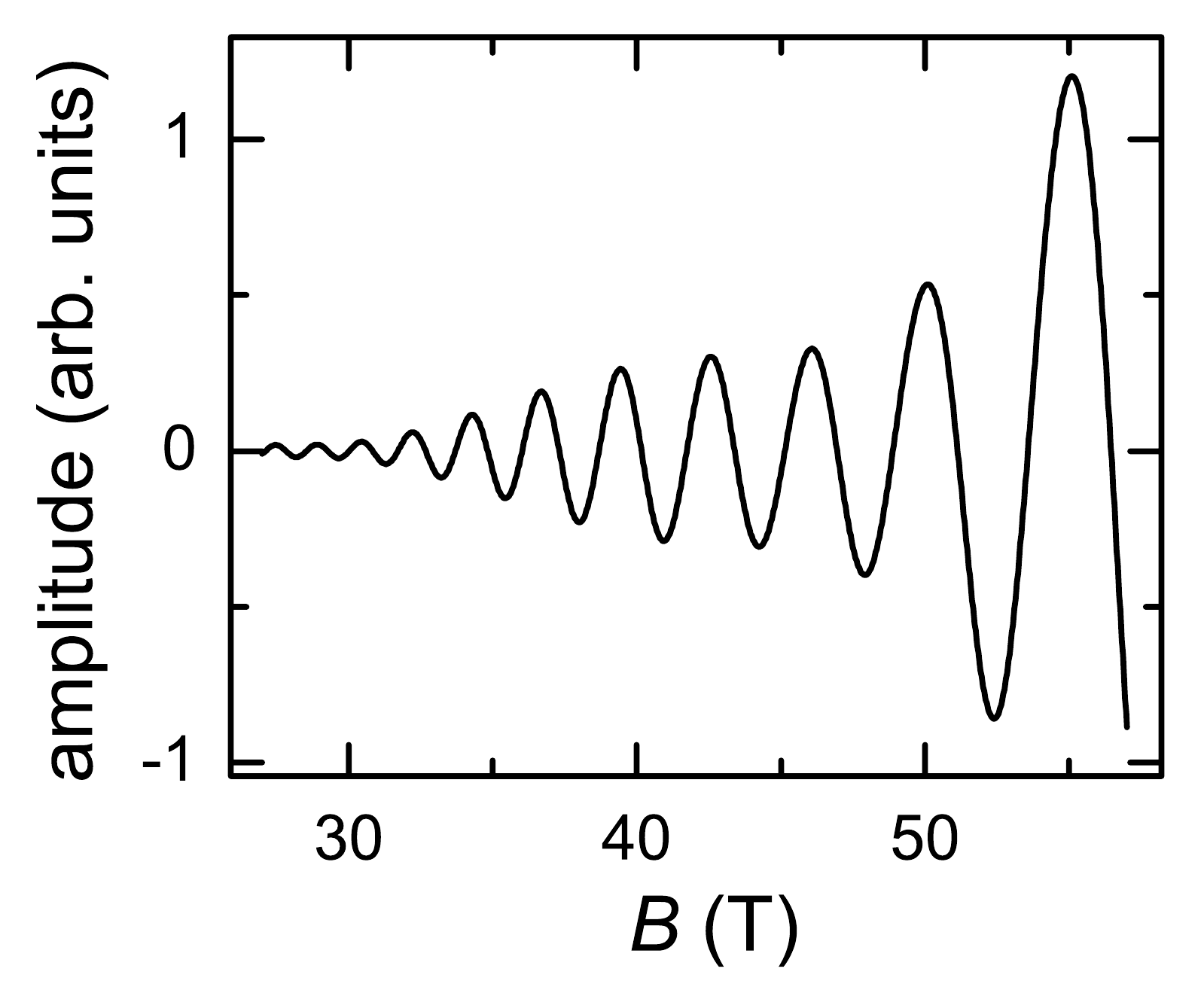}
\caption{Simulation of the magnetization (or magnetic torque $\tau\propto-M_zB$) from Ref.~\cite{audouard1} using Eqn.~(\ref{magnetization}) and four adjustable parameters $F_0=(\hbar/2\pi e)A_0\approx$~513~T, $\Delta F=(\hbar/2\pi e)A_1\approx$~40~T and $\Delta F_2=(\hbar/2\pi e)A_2\approx$~-10~T in Eqn. (\ref{area}) and $\Gamma=$~170~T in Eqn.~(\ref{prefactors}). These values yield $F_{\rm neck}\approx$~463~T and $F_{\rm belly}\approx$~543~T. %
We have assumed the phase to be a fixed quantity in Eqn.~(\ref{magnetization})~\cite{note1}. We have also assumed 
$T=$~1~K, $m^\ast/m_{\rm e}=$~1.7 and $g=$~0 to be fixed quantities, although inaccuracies in $T$ or $m^\ast$ are easily compensated by an adjustment of $\Gamma$ and thus have little affect on the waveform.}
\label{proust}
\end{figure}

We now turn to $\theta$-dependent simulations of $M_z$ in underdoped YBa$_2$Cu$_3$O$_{6+x}$ for several different Fermi surface topology scenarios. Recent quantum oscillation measurements on YBa$_2$Cu$_3$O$_{6.54}$~\cite{audouard1,sebastian2} (detected over a wide interval in magnetic field $B\approx\mu_0H$) provide estimates for the relevant parameters. With the exception of the weak 637~T frequency~\cite{audouard1}, the full waveform from Ref.~\cite{audouard1} can be reproduced in Fig.~\ref{proust} by numerical integration of Eqn.~(\ref{magnetization}) and the adjustment of only four parameters~\cite{note1} (listed in the figure caption). 

\begin{figure}[htbp!]
\centering
\includegraphics[width=0.45\textwidth]{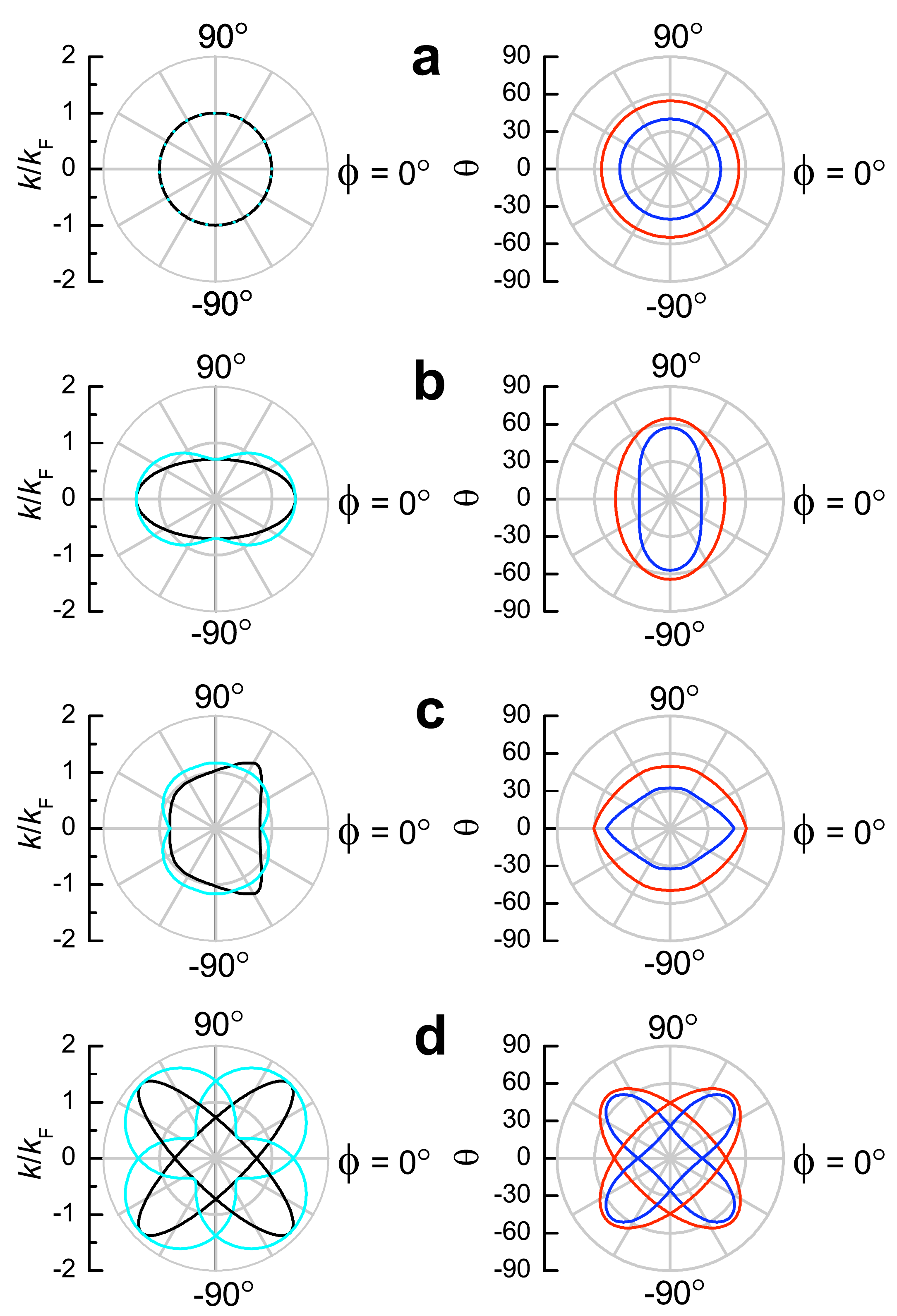}
\caption{Polar plots (cyan) of $k_\|$ versus $\phi$ (left-hand-side) for different cylinder cross-sections (black), together with polar plots of $\theta_1$ (blue) and $\theta_0$ (red) versus $\phi$ (right-hand-side). The estimates of $\theta_1$ and $\theta_0$ are made at $B=$~40~T, corresponding to $\Delta F/B=$~1 in Eqn.~(\ref{nodes}) and Fig.~\ref{nodeplot}. Polar plot {\bf a} corresponds to a simple circular corrugated cylinder cross-section for which $k_\|=k_{\rm F}$, {\bf b} corresponds an elliptical electron pocket with an ellipticity of $\eta=$~2 and orientation $\zeta=$~0 predicted in $\frac{1}{8}^{\rm th}$ stripe models that combine spin and charge ordering, while {\bf c} corresponds to an electron pocket of unusual cross-section predicted by certain d-density wave or spiral ordering models. {\bf d} corresponds to highly elliptic hole pockets ($\eta=$~3.5) for which two different orientations $\zeta\approx\pm$~45$^\circ$ within the $k_x$, $k_y$ plane are expected, giving rise to two different values of $k_\|$ for each value of $\phi$.}
\label{polar}
\end{figure}

\begin{figure}[htbp!]
\centering
\includegraphics[width=0.45\textwidth]{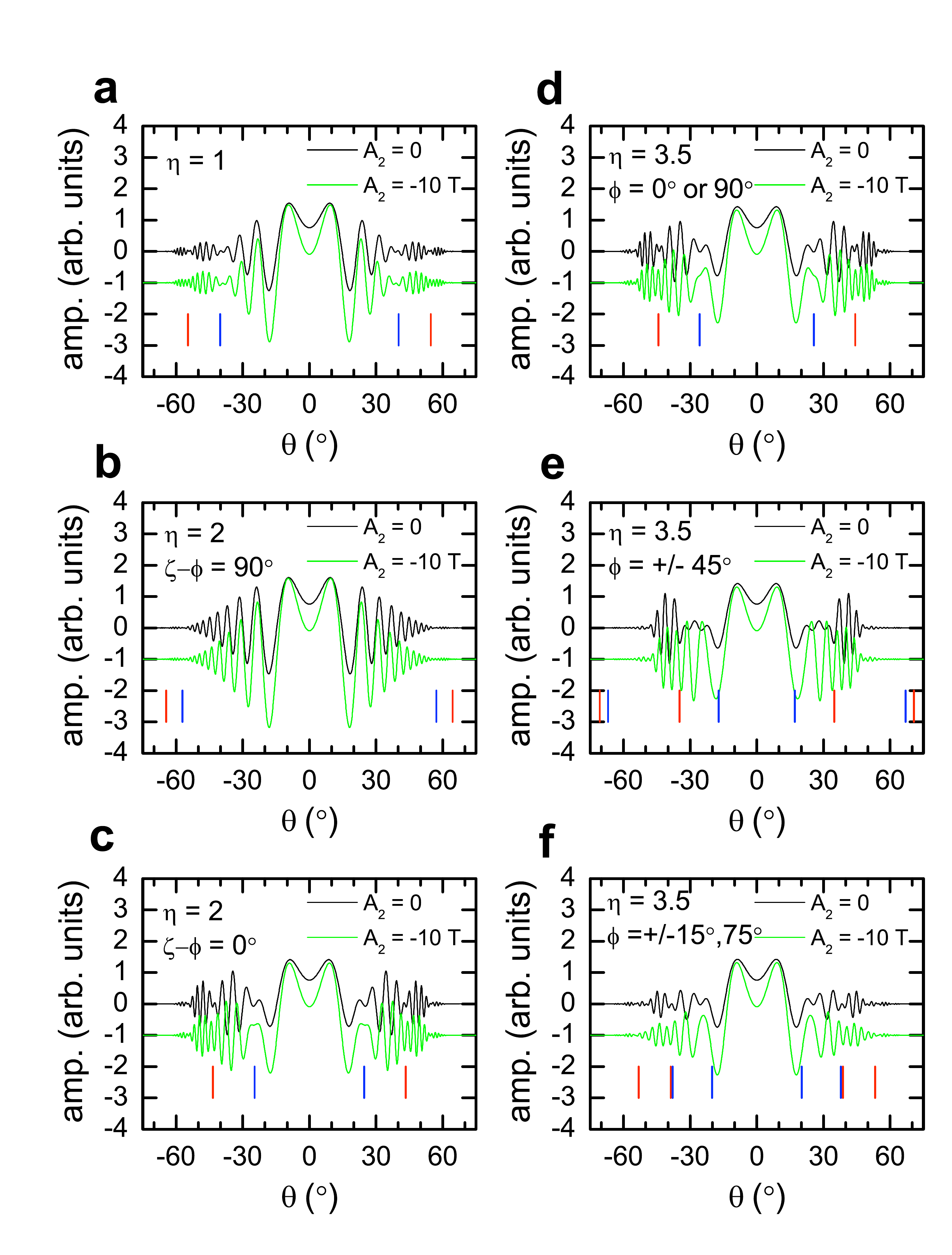}
\caption{Examples of $\theta$-dependent simulations at $B=$~40~T made using Equation (\ref{magnetization}) using the parameters listed in Fig.~\ref{proust} caption (green curves) and using Equation (\ref{magnetization2}) in which $A_2$ is neglected (black curves) for several different scenarios depicted in Fig.~\ref{polar}. Curve {\bf a} is the expected $\theta$-dependence for a simple circular cross-section, while {\bf b} and {\bf c} are the expected curves obtained on rotating about the major and minor axes (respectively) of the elliptical pocket with $\eta=$~2 shown in Fig.~\ref{polar}b. {\bf d}, {\bf e} and {\bf f} are examples of the $\theta$-dependence expected for the highly elliptical hole pockets in Fig.~\ref{polar}d in which there are two different values of $k_\|$ for each value of $\phi$ (with the exception of $\phi=$~0 or 90$^\circ$ ({\bf d}) where they become degenerate). The red and blue vertical lines point to the locations of $\theta_0$ and $\theta_1$ respectively.}
\label{anglesweeps}
\end{figure}

We begin by considering scenarios in which the $F\sim$~500~T frequency corresponds to an electron pocket cylinder, for which a single value of $k_\|$ is expected at each azimuthal angle $\phi$. Several different proposals for the groundstate of YBa$_2$Cu$_3$O$_{6+x}$ predict electron pockets that are approximately circular in cross-section and situated near ${\bf k}=(\pi,0)$ in Fig.~\ref{schematic}a. These include ${\bf Q}=(\pi,\pi)$ spin- or d-density wave ordering~\cite{dimov1,sebastian1}, certain types of spin stripe ordering in which ${\bf Q}=(\pi[1\pm2\delta],\pi)$ where $\delta=\frac{1}{8}$~\cite{millis1} and helical or spiral ordering~\cite{sebastian1,dimov1}. Strong coupling treatments in which the hopping of quasiparticles is frustrated by antiferromagetic correlations can also predict circular electron pockets under certain conditions~\cite{hozoi1}, although in this case they are situated at ${\bf k}=(\pi,\pi)$. In all these cases, a circular electron pocket implies that $k_\|\approx k_{\rm F}$ for all $\phi$, in which case nodes corresponding to $m=$~1 and 0 should be observed at $\theta_1\approx$~40$^\circ$ and  $\theta_0\approx$~55$^\circ$ when $B=$~40 T at all values of $\phi$ (as shown in Fig.~\ref{polar}a). Figure~\ref{anglesweeps}a shows the calculated $\theta$-dependent magnetization (using the same parameters as in Fig.~\ref{proust}). Nodes at  $\theta_1$ and $\theta_0$ are clearly evident in the simulation made using Eqn.~(\ref{magnetization3}) in which the second harmonic of the corrugation $A_2$ is neglected (black curve), and remain discernible and at the same positions in the simulation made using Eqn.~(\ref{magnetization}) in which $A_2$ is included (green curve).

Under certain conditions, the electron cross-sectional area can deviate from an ideal circular form. In $\delta=\frac{1}{8}$ stripe models that incorporate both spin and charge ordering (e.g. Figs. 3c and 3d from Ref.~\cite{millis1}), for example, the cross-section is expected to become elongated along $k_x$. In Fig.~\ref{polar}b this situation is approximated by an ellipsoidal pocket with $\eta=$~2 and $\zeta=$~0. The reduced symmetry of the pocket is reflected in the $\phi$-dependence of the angular sweeps in Fig.~\ref{anglesweeps}a and b. For $\phi=$~90$^\circ$ (in which, since $\zeta=$~0, the axis of rotation corresponds to the major axis) $\theta_1$ and $\theta_0$ are pushed out to very high angles in Fig.~\ref{anglesweeps}b, becoming practically unobservable for the simulation made using Eqn.~(\ref{magnetization}) in which $A_2$ is included. For $\phi=$~0, by contrast, $\theta_1$ and $\theta_0$ are reduced to small angles in Figs~\ref{polar}b and \ref{anglesweeps}c. If the unreconstructed Fermi surface is more circular in form~\cite{dimov1}, spiral or d-density wave order can give rise to trapezoidal-shaped electron pocket cross-sections, as shown in Fig.~\ref{polar}c (using ${\bf Q}=(\pi[1+2\delta],\pi)$ where $\delta=$~0.1. In spite of the loss of rotational symmetry, the $\phi$-dependent $k_\|$ is similar to that of an ellipse, but with its major axis along $k_y$. Since the $\phi$ dependences of $\theta_1$ and $\theta_2$ are rotated by 90$^\circ$ with respect to Fig.~\ref{polar}b, their measurement can by used to distinguish stripe and spiral or d-density wave scenarios, or perhaps other scenarios~\cite{harrison1}

Turning now to the case where the $F\sim$~500~T frequency originates from hole pockets~\cite{lee1,hossain1,doiron1}, the $\phi$-dependences of $k_\|$, $\theta_1$ and $\theta_0$ in Fig.~\ref{polar}d and the forms of the $\theta$-dependent curves in Figs.~\ref{anglesweeps}d-f become very different from those of electron pockets. In the hole pocket simulations we have assumed $\eta=$~3.5, which is approximately what might be expected were the Fermi surface obtained in recent angle-resolved photoemission experiments~\cite{hossain1} folded about the $(\pi,0)-(0,\pi)$ line. Since there are two different orientations $\zeta\approx\pm$~45$^\circ$ to be counted in Fig.~\ref{polar}d and Eqn.~(\ref{magnetization}), two different values of $k_\|$ give rise to four simultaneously beating frequencies, leading to more complicated $\theta$-dependent curves in Figs.~\ref{anglesweeps}d-f. On the other hand, the $\theta$-dependent curves exhibit a distinctive fourfold symmetry that changes rapidly with $\phi$ owing to the highly elliptical cross-sections of the pockets. For most arbitrary values of $\phi$ (e.g. Fig.~\ref{anglesweeps}e), the contribution from two very different values of $k_\|$ make the nodes somewhat difficult to locate. Distinct nodes do appear, however, at $\phi=$~0 and 90$^\circ$ in Fig.~\ref{anglesweeps}d when the two values of $k_\|$ become degenerate, or near $\phi\approx\pm$~15$^\circ$ and $\pm$~75$^\circ$ in Fig.~\ref{anglesweeps}f due to a serendipitous degeneracy of $\theta_1$ and $\theta_0$. For this reason, rather detailed $\phi$-dependent measurements are required to pin down the precise geometry of the cross-section of hole pockets.

In summary, we have shown that nodes in the quantum oscillation amplitude detected in dual axis rotation experiments can be used to infer the cross-sectional shape of corrugated Fermi surface cylinders in underdoped YBa$_2$Cu$_3$O$_{6+x}$, which can help to answer the question as to whether the prominent $\sim$~500~T frequency originates from electron- or hole-like carriers, as well as providing essential clues for the groundstate description. An advantage of this proposed method over AMRO experiments~\cite{yamaji1,nam1} is that the shape of the cross-section is linked directly to the pocket responsible for the quantum oscillations (which needs always to be assumed in AMRO experimenta). A further advantage is that the predicted nodes occur at small values of $\theta$ where $H_{\rm c2}$ is lower, thereby proving less of an obstacle for Fermi surface measurements. 

There are, however, potential caveats. In the present simulations we have assumed that the corrugation of the Fermi cylinders is caused by simple interlayer hopping terms $t_{c,n}$. More complicated forms of warping that vary with $k_x$ and $k_y$ may lead to some degree of variation from the present simulations. A $k_z$-dependent damping factor, meanwhile (for which there is presently no evidence~\cite{note1}), will diminish the presence of nodes, while Zeeman splitting effects (presently neglected by setting $g=0$~\cite{sebastian1}) will lead to additional nodes unrelated to the corrugation. Nodes due to Zeeman splitting (often called `spin splitting zeroes') typically depend only on $\theta$, in contrast to those caused by corrugation described here, which depend on $B$, $\phi$ and $\theta$.


\begin{thebibliography}{99}
\bibitem{note1} The ability of Eqn.~(\ref{magnetization}) to reproduce the waveform with only four parameters suggests that a uniform $\Gamma$ (which is proportional to the quasiparticle scattering rate~\cite{shoenberg1}) provides an adequate approximation, but that a $n=$~2 component of corrugation must be taken into consideration. The ability of Eqn.~(\ref{magnetization}) to reproduce the experimental magnetization implies that there is no Berry phase. The sign of $\tau$ (which results from a vector product in the experiment) is assumed to be negative in order to correctly account for the inferred sign of $M_z$ from Ref.~\cite{jaudet1}.
\end{thebibliography}
\end{document}